# SAPIENS CHAIN: A BLOCKCHAIN-BASED CYBERSECURITY FRAMEWORK

Yu Han[1], Zhongru Wang[1,2] （✉）, Qiang Ruan[3] and Binxing Fang[1]

[1]Key Laboratory of Trustworthy Distributed Computing and Service (BUPT),
Ministry of Education, Beijing, China
{hany,wangzhongru,fangbx}@bupt.edu.cn
[2]Zhejiang Lab, Hangzhou, China
[3]Beijing DigApis Technology Co., Ltd, Beijing, China
ruanqiang@digapis.cn

## ABSTRACT

*Recently, cybersecurity becomes more and more important due to the rapid development of Internet. However, existing methods are in reality highly sensitive to attacks and are far more vulnerable than expected, as they are lack of trustable measures. In this paper, to address the aforementioned problems, we propose a blockchain-based cybersecurity framework, termed as Sapiens Chain, which can protect the privacy of the anonymous users and ensure that the transactions are immutable by providing decentralized and trustable services. Integrating semantic analysis, symbolic execution, and routing learning methods into intelligent auditing, this framework can achieve good accuracy for detecting hidden vulnerabilities. In addition, a revenue incentive mechanism, which aims to donate participants, is built. The practical results demonstrate the effectiveness of the proposed framework.*

## 1. INTRODUCTION

In recent years, the applications of Internet of Things, Internet of Vehicles, and Mobile Payment have been more and more popular and deeply affect human life [1][2]. However, these applications face more serious security risks than before [3]. For example, more than 70 countries and regions were attacked by the newly produced computer virus WannaCrypt0r 2.0 and suffered high damages [4]. Uber lost large scales of sensitive information, which may be related to 57 million users and 7 million drivers [5]. Besides the traditional security problems, new techniques, such as blockchain, may become exposed to security threats. For example, the famous incident DAO occurred in Ethereum and the attackers stole about 3.5 million Ether, which was worth about 60 million dollars at that time, owing to a smart contract vulnerability [5][6]. The high yield of successful attacks drives the "prosperity" of the black industry.

To deal with the aforementioned cybersecurity problems, many studies are proposed [7-10]. Not surprisingly, existing methods mostly focus on centralized models and have the following drawbacks. First, it's difficult to manage data storage and security dynamically. Traditional data storage and security management are always built in the trust and centralized environment, while attacks on the central management nodes may devastate private data and the networks [11]. Second, it's hard to cope with the high-intensity attacks timely with limited resources. In addition, the participants require a security interactive platform, which can protect their privacy and avoid information leakage. Third, the white hat hackers can only obtain little revenue from the security vendors, such that they have low interests in helping vendors fix their vulnerabilities.

To tackle these challenges, we design a blockchain-based framework, named by Sapiens Chain, that protects all participants by using a decentralized, non-monopoly and non-intermediate model. We make the following contributions in this work.

First, we design a smart contract for all participants, where the transactions are written into blocks and almost impossible to be modified. By defining the incentive mechanism on smart contracts, the Proof-Of-Cnocept (POC) providers can be awarded if the task result is adopted by the framework. The task details and identities of participants will be disclosed, such that the privacy of participants is guaranteed.

Second, we introduce two kinds of nodes, including the ordinary nodes and the fog nodes. Ordinary nodes perform task assignment, vulnerability detection, POC construction, and POC auditing, while the fog nodes perform node scheduling and storage for POCs and vulnerabilities. For reducing the computational resource overhead as much as possible, we propose a novel node scheduling method, which combines the proof of work with the distances between nodes.

Third, we propose a novel model that can audit websites, applications and smart contracts automatically. For websites, the model can automatically identify network assets and vulnerabilities through knowledge graphs and association rules. For applications and smart contracts, the model first extracts basic semantic information through dependency graphs, and then discover vulnerabilities within the codes by performing analysis on the semantic information.

The rest of the paper is organized as follows. We review the related work in Section 2 and propose the framework in Section 3. Section 4 introduces roles, techniques and operational modes of the framework. We introduce the typical application in Section 5 and conclude the paper in Section 6.

## 2. RELATED WORK

In this section, we review some related work, including the existing blockchain-based cybersecurity protection methods and systems.

### 2.1. Blockchain-based Cybersecurity Studies

Many novel cybersecurity techniques have been used in website security [12] [13], application security [14] and blockchain security [15]. For example, Nikolic et al. [16] present MAIAN, the first tool for precisely specifying and reasoning about trace properties, which employs inter-procedural symbolic analysis and concrete validator for exhibiting exploits. Tsankoc et al. [8] present Securify, a security analyzer for Ethereum smart contracts that is scalable, and able to prove contract behaviors as safe/unsafe with respect to a given property.

Recently, blockchain technology has made significant contributions to cybersecurity due to its immutability, traceability, decentralization, and transparency [12-17]. Zyskind et al. [18] propose to protect application data using blockchain, which separates data from permissions, records permission settings and data access in blockchain, enabling full control of data access permissions and transparent access procedures. Azaria et al. [19] propose a medical data management model based on blockchain and smart contract, which records data permissions and operations in the blockchain, and is executed by smart contracts to implement data authentication, confidentiality, auditing, and sharing. Buldas et al. [20] propose a blockchain-based keyless signature framework, which records the root hash value in the chain and performs multi-file signature, which increases the overhead of falsifying signature files, ensuring the integrity of the file. Ali et al. [21] propose a distributed domain name resolution system based on blockchain, where this system can effectively resist DDoS attacks by layering the domain name resolution logic and the underlying consensus mechanism.

However, previous approaches focus on only one or two aspects of cybersecurity, which cannot provide a fair and trust environment. Our framework can not only detect vulnerabilities, but also protects all participants' privacy by using a decentralized, non-monopoly and non-intermediate model.

## 2.2. Blockchain-based Cybersecurity Applications

Several applications, such as CertiK [22], SECC [23] and DVP [24], have been produced to protect cybersecurity. CertiK uses formal verification techniques to transform smart contracts into mathematical models and validate models through logical calculus to prove the security. The automated auditing tools are deployed on the server and improve capabilities via plug-in provided by white hats. In order to solve the inherent vulnerabilities of the original public chain, wallet, and transactions, SECC essentially recreates a public chain, the nodes on which are safe nodes and the applications on which are safe applications. DVP built a vulnerability platform on blockchain to ensure its security. The system operates on the blockchain and provides the required power based on the distributed network of participants, who use the agreement points to pay, receive or enhance the verification service.

In contrast to the previous framework, Sapiens Chain can detect the vulnerabilities automatically and can handle website security, application security, and blockchain security simultaneously.

## 3. THE ARCHITECTURE OF THE FRAMEWORK

In this section, we first propose the overview of the proposed framework and then introduce the structure.

The overview of the proposed framework is shown in Figure 1. The computing nodes of the Sapiens Chain are decentralized and thus each node won't be affected by others. The users submit their tasks including website tasks, application tasks, and smart contract tasks through the browser, the fog nodes in Sapiens Chain first distinguish the type of the task, and then segment tasks into several parts, running the algorithms to select proper nodes to deal with the task, and finally gather the results into a report.

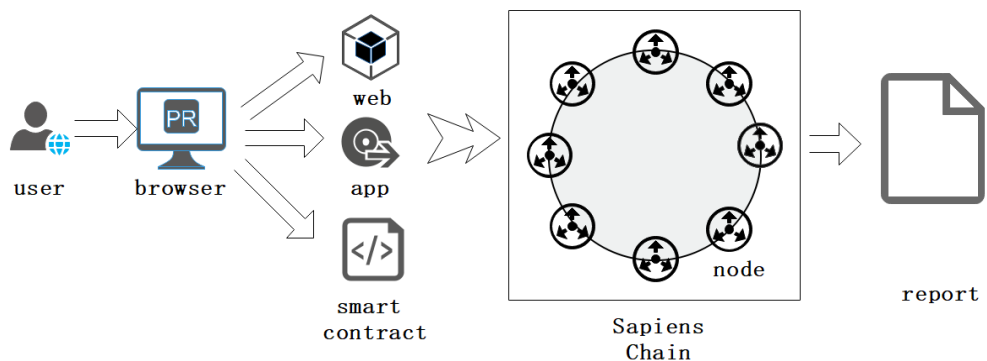

Figure 1. Sapiens Chain overview

The hierarchical structure of Sapiens Chain is shown as Figure 2. Sapiens Chain has 4 layers, including the resource layer, the transport layer, the contract layer, and the application layer.

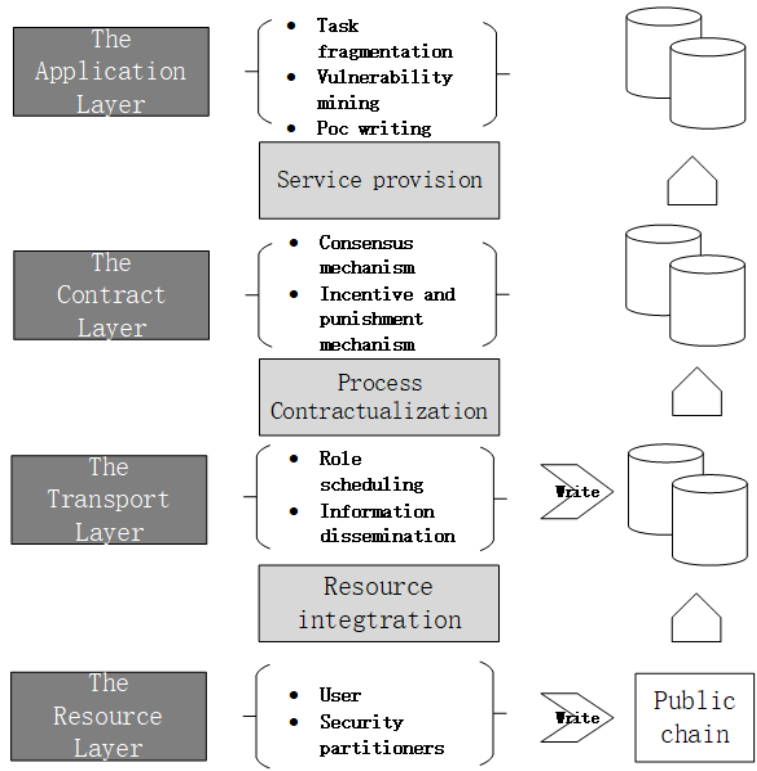

Figure 2. The architecture of Sapiens Chain layers

a) The resource layer integrates a large number of resources, which ensures that Sapiens Chain can have a large detection capacity. The essence of the resource layer is a distributed ledger that records the service information of all participants [25]. The record format is related to the public chain where the framework runs, and the records are guaranteed to protect the rights of the service provider from being infringed.

b) The transport layer, which relies on P2P network, is used to transmit information, which includes resource scheduling and information dissemination [26]. The nodes in Sapiens Chain are theoretically accessible to each other. The ledgers of nodes will be synchronized after finishing tasks.

c) The contract layer is designed to contract the service process, including both consensus and incentive and punishment mechanisms. The consensus mechanism is consistent with the consensus mechanism of the underlying blockchain, and all kinds of nodes are ranked according to the active level and task completion degree [27]. According to the accuracy and fairness of the results of the consensus mechanism, this framework can achieve good audit performance. Based on the ruling result of the consensus mechanism, the incentive and punishment mechanism rewards or punishes nodes and writes the results of arbitration on the chain and then can protect the rights of the contributors.

d) The application layer realizes the fragmentation and classification of tasks, and the centralized security service of traditional security vendors is distracted, distributed to each node, therefore the benefits are shared [28]. Each node accesses the service through the blockchain browser at the application layer. The distribution of the fragmentation task adopts a redundancy mechanism. The selected nodes get several non-repeating segments, and each segment is dispersed to several non-repeating nodes. The nodes perform three different types of tasks, which includes manual vulnerability detecting, POC writing and auditing, and running automatic auditing tools.

## 4. THE OPERATION MODE OF THE FRAMEWORK

In this section, we first introduce two kinds of nodes and the inherent techniques, and then propose how to operate this framework.

### 4.1. Nodes

Sapiens Chain includes two kinds of nodes: ordinary nodes and fog nodes.

#### 4.1.1. Ordinary Nodes

The ordinary node contains 6 different roles, which is shown as follows.

• User. The user submits the tasks, which requires to be detected by Sapiens Chain. The privacy of users and the immutability of the transactions can be protected by smart contracts defined between users and Sapiens Chain.

• Proof of Concept Developer (POCD). The POCD provides POCs. Essentially, each node of Sapiens Chain can be a POCD, whose gains are related to the number of accepted POCs and called POCs.

• Proof of Concept Auditor (POCA). The POCA audits the POCs, which is provided by POCD. The POCAs are generated through an arbitration mechanism among POCDs.

• White Hat Hacker (WHH). The WHH is the provider that can supply manual audit services. WHHs are selected through a scheduling scheme, which is designed by Sapiens Chain.

• White Hat Auditor (WHA). The WHA is one of the WHHs, which has a higher active degree and generated by an arbitration mechanism. Its task is to test the auditing results submitted by WHHs.

• Computational Resources Owner (CRO). The CRO deploys an automatic vulnerability audit tool and is called by a scheduling algorithm. As providing automatic auditing, CROs benefit from their own computational resources, which can reduce the operating costs and increase the enthusiasm simultaneously.

#### 4.1.2. Fog Nodes

The fog node is responsible for node scheduling, POC storage, report storage, and key assignment for each ordinary node [29]. It runs a proof of work and node distance-based scheduling mechanism that can select appropriate CROs to perform the detection service. In addition, the fog nodes can store the submitted vulnerabilities, accepted POCs and test reports, such that these relevant resources can be reused.

Figure 3 shows how different roles of nodes interact with each other. The user issues a task to the fog node, and the fog node schedules CROs, WHHs or WHAs according to the given tasks. CROs receive POCs which are provided by POCDs and audited by POCAs and then run detection processes. The vulnerabilities submitted by WHHs will be audited by WHAs and finally received by the fog nodes, while WHHs can also submit vulnerabilities to users through the fog nodes. The fog nodes can distribute tasks and collect reports.

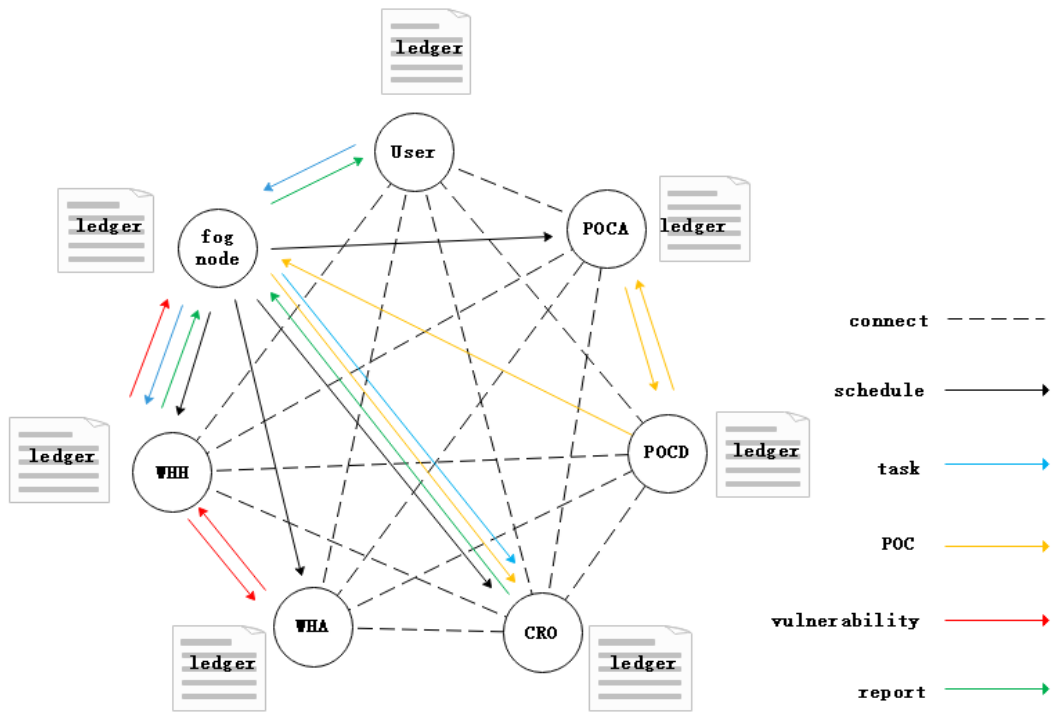

Figure 3. Roles in Sapiens Chain

## 4.2. The Algorithms

The algorithms of Sapiens Chain can be categorized into scheduling algorithms and security algorithms.

The scheduling algorithms called by the fog nodes, are used to allocate tasks or computational resources dynamically according to the different roles of ordinary nodes. These algorithms run for POCD, POCA, CRO, WHH, and WHA selection. Following the principle of proximity, the algorithms rank the computational resources according to the proof of work mechanism and select the nodes closest to the users [30]. Since task auditing by WHHs requires professional knowledge, we ensure the ability of the nodes by applying the incentive mechanism and punishment mechanism which will be introduced later.

The security algorithms use symbolic execution, taint analysis, and formal verification to detect website, application and smart contract vulnerabilities. For websites, Sapiens Chain can detect vulnerabilities with thousands of built-in predefined knowledge graphs and association rules, which can automatically identify asset information and then realize deep detection of security vulnerability. For applications and smart contracts, Sapiens Chain builds the dependency graphs to extract basic semantic information. With these semantics and the control flow graphs, it can match vulnerability patterns, which can avoid possible path execution errors and improve the detection accuracy.

## 4.3. Work Mode

### 4.3.1. Reward Mode

The reward mode refers to that the users submit tasks and then select the automatic audit service or the manual audit service. For the automatic audit service, the CROs are scheduled by the fog nodes to call the automatic detection tools. POCs stored in fog nodes will be installed on the tools on CRO, and then CRO runs the tools to output a report about vulnerabilities and patches. For the manual audit mode, the fog nodes schedule WHHs, who can also submit reports to WHAs. This process ends until the reports are approved.

### 4.3.2. Claim Mode

The claim mode refers to the mode where the WHHs actively submit the vulnerabilities, which can be claimed by users if needed. The WHHs can encrypt the details of vulnerabilities with users' public keys and send them to the fog nodes for storage. Then the fog nodes transport the information to users and determine whether they will claim. The WHHs can benefit if the vulnerabilities are claimed.

### 4.4. Incentive and Punishment Mechanism

#### 4.4.1. Incentive Mechanism

In order to avoid issues of network abuse and encourage more nodes to provide computational resources, we propose the fuel called SACF, which can be exchanged in Sapiens Chain. The SACF can be regarded as a reward after each role provides effective services, and it can be paid for services purchasing by users. Different roles can be rewarded under the following situations. (1) POCDs submit POCs and the POCs are adopted after review; (2) POCAs participates in POC audit and their final audit results are adopted, (3) WHHs submit vulnerabilities which are also adopted, (4) WHAs audit the vulnerabilities and the results are adopted, (5) CROs provide complete audit services and finally output an available audit report.

#### 4.4.2. Punishment Mechanism

In order to deal with the problem of node dishonesty, we propose a punishment mechanism. For CRO, we define a parameter that measures the average processing capacity, which corresponds to the number of running tasks and computational resources. The parameter will drop when the CRO did not complete the task. When this parameter of the nodes becomes zero, we will abandon such nodes. For WHHs, if the vulnerabilities submitted in the reward mode are not approved or these submitted in the claim mode are not successfully claimed, we will punish them and decrease their rankings which denote the priority to be scheduled. We omit the mechanisms for POCAs and WHAs as it's similar to the WHH case.

## 5. PRACTICAL RESULTS AND TYPICAL APPLICATIONS

In this section, we first introduce the practical results of the proposed framework and then propose typical applications.

### 5.1. Practical Results

We implement our framework as a security auditing platform, which aims to detect risks and vulnerabilities and give suggestions for improvement. In our framework, each selected CRO node is an automatic audit tool, which can audit websites, applications and smart contracts. Taking several test websites for example, we run the CRO node and get the result as below.

As shown in Figure 4, 80% of detected websites are at low risk level, while 20% are at high level. This demonstrates that, our framework can distinguish risk levels. As shown in Figure 5, in one test website, we can detect 368 vulnerabilities, and 179 vulnerabilities are at high risk level. This demonstrates that our framework is effective in detecting vulnerabilities. After detection, the node will output a report containing specific vulnerabilities and corresponding suggestions.

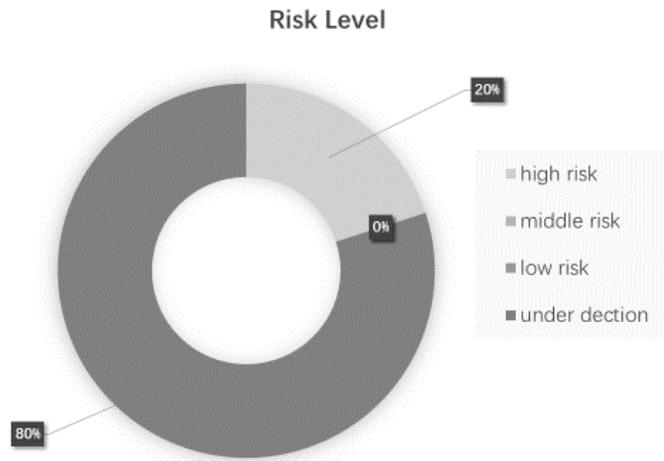

Figure 4.The risk types of website

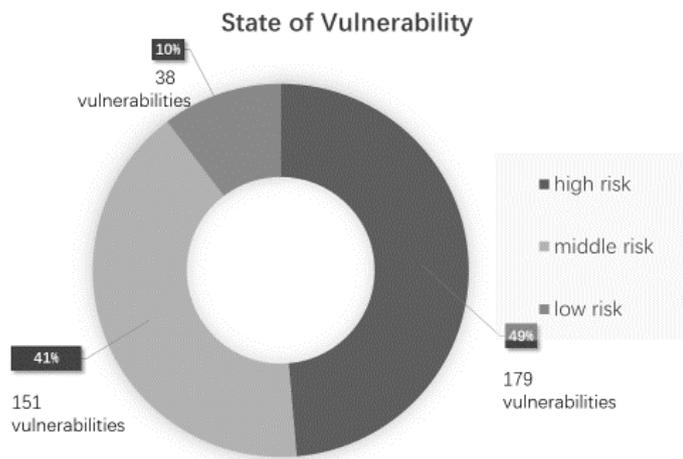

Figure 5. The risk types of vulnerabilities

## 5.2. Typical Applications

The framework has the following application scenarios.

(1) Website, application and smart contract security. In order to protect the website from attacks, security vendors often exploit and repair vulnerabilities with the help of the security team or the vulnerability platform. However, the security teams of different vendors are not strong enough to cope with the surge in cybersecurity, such that they may not identify hidden vulnerabilities. Sapiens Chain builds a trust security audit platform for all practitioners in the field of cybersecurity so that we can provide vulnerability audit services with high accuracy. During the auditing process, Sapiens Chain can protect the privacy and reward according to the donation, which can attract more and more white hats.

(2) Shared Economy. In Sapiens Chain, CROs can earn revenue by sharing idle network bandwidth, storage space, and computational resources. Through the trusted interactive network built by blockchain technology, Sapiens Chain allocates resources through the scheduling algorithms, which can closely meet the requirement of the sharing economy.

## 6. CONCLUSIONS

In this paper, we proposed Sapiens Chain, a blockchain-based cybersecurity framework for security detection and protection. Sapiens Chain leverages the combination of blockchain technology and artificial intelligence that distribute computational resources and accomplish tasks automatically. Based on blockchain, the framework collect resources for the missions, at the same time using the distributed ledger to guarantee the immutability of the reward process. Using artificial intelligence, each CRO node is an automated audit tool, and its audit capacities can be continuously improved through machine learning whose samples come from the manual detection process, which means the samples are endless. The practical results demonstrate the effectiveness of the proposed framework.